%% file: StretchedFG.tex
\documentclass[journal]{IEEEtran}

\usepackage{graphicx,multicol,subfigure}
\usepackage{psfrag,amsbsy,pstricks,amsmath,amsfonts,amssymb,cite}
\usepackage{algorithm,algpseudocode}
\usepackage{epstopdf}

\usepackage{color}
\usepackage{amsmath}

\usepackage{booktabs}

\input{mydefinitions.tex}

\begin{document}
%
\title{A New Combination of Message Passing Techniques for Receiver Design in MIMO-OFDM Systems}

\author{Chuanzong~Zhang, Zhengdao~Yuan, Zhongyong Wang and Qinghua Guo 
\thanks{This work is supported by the National Natural Science
Foundation of China (NSFC 61172086, NSFC U1204607, NSFC 61201251). 
}
\thanks{C. Zhang is with the School of Information Engineering, Zhengzhou University, Zhengzhou 450001, China, and the Department of Electronic Systems, Aalborg University, Aalborg 9220, Denmark (e-mail: ieczzhang@gmail.com).}
\thanks{Z. Yuan is with the National Digital Switching System Engineering and Technological Research and Development Center, and the Zhengzhou Institute of Information Science and Technology, Zhengzhou 450001, China (e-mail: yuan\_zhengdao@foxmail.com).
}
\thanks{Z. Wang is with the School of Information Engineering, Zhengzhou University, Zhengzhou 450001, China (e-mail: iezywang@zzu.edu.cn).}
\thanks{Q. Guo is with the School of Electrical, Computer and Telecommunications Engineering, University of Wollongong, Wollongong, NSW 2522, Australia, and also with the School of Electrical, Electronic and Computer Engineering, University of Western Australia, Crawley, WA 6009, Australia (e-mail: qguo@uow.edu.au).}
}
\maketitle

\begin{abstract}
In this paper, we propose a new combined message passing algorithm which allows belief propagation (BP) and mean filed (MF) applied on a same factor node, so that MF can be applied to hard constraint factors. Based on the proposed message passing algorithm, a iterative receiver is designed for MIMO-OFDM systems. Both BP and MF are exploited to deal with the hard constraint factor nodes involving the multiplication of channel coefficients and data symbols to reduce the complexity of the only BP used. The numerical results show that the BER performance of the proposed low complexity receiver closely approach that of the  state-of-the-art receiver, where only BP is used to handled the hard constraint factors, in the high SNRs.
\end{abstract}


\IEEEpeerreviewmaketitle

\section{Introduction}\label{Sec:intro}
\IEEEPARstart{R}{ecently}, multiple-input multiple-output orthogonal frequency division multiplexing (MIMO-OFDM) is a key technology
for many wireless communication standards, due to its high spectral efficiency \cite{Stuber2004}. Message passing techniques performing
Bayesian inference on factor graphs~\cite{Wainwright2008} have proven to be a very useful tool to design receivers in communication systems. And there are several message passing based receivers~\cite{Navarro2011,Wu2016,Wen2016,Daniel2016,Yuan2017} with joint channel estimation and decoding for MIMO-OFDM systems in literature.

Belief propagation (BP), also known as sum-product algorithm~\cite{Kschischang2001}, is the most popular message passing technique  for its excellent performance, especially applied in discrete probabilistic models. BP has been widely used to design iterative  receivers in digital communications. Its remarkable performance, especially when applied to discrete probabilistic models, justifies its popularity. However, its complexity may become intractable in certain application contexts, e.g. when the probabilistic model includes both discrete and continuous random variables.
As an alternative to BP, variational methods based on the mean field (MF) approximation have been initially used in quantum and statistical physics. The MF approximation has also been formulated as a message passing algorithm, referred to as variational message passing (VMP) algorithm~\cite{Winn2005}. It has primarily been used on continuous probabilistic conjugate-exponential models.

Yedidia et. al. proposed  to derive the fixed point equations of both BP and MF approximation by minimizing region-based free energy approximation in \cite{Yedidia2005}.
A unified message passing framework which combines BP with MF~\cite{Riegler2013} is proposed based on a particular region-based free energy approximation. Combined BP-MF  allows that one divides the factor nodes on a factor graph into two disjoint subsets: a BP part and a MF part. The messages passed to or outgoing from a factor node in the BP
part are computed by BP rule, while MF rule for a factor node in the MF part. Therefore, it keeps the virtues of BP and MF but avoid their respective drawbacks. The MIMO-OFDM receivers~\cite{Navarro2011,Daniel2016,Yuan2017} are proposed using combined BP-MF.

In this paper, we heuristically apply both BP- and MF-like rule for a same factor node when designing  message passing receiver for MIMO-OFDM systems. Generally speaking, MF rule is not suitable for hard constraint factor nodes for the logarithm operation to the factor. We handle a special kind of hard factor nodes on a stretched factor graph representation~\cite{Yuan2017} using BP-like rule, yielding an exponential function, and then MF-like rule becomes possible to be applied. A new message passing receiver with low complexity is obtained the hybrid message computation of a same factor node.

\textit{Notation}- Boldface lowercase and uppercase letters denote vectors and matrices, respectively.
The expectation operator with respect to a pdf $g(x)$ is expressed by $\left\langle f(x) \right\rangle_{g(x)} = \int f(x) g(x) dx / \int g(x') dx' $, while $\textrm{var}[x]_{g(x)}= \left\langle \vert x\vert^2\right\rangle_{g(x)}-\vert\left\langle x\right\rangle_{g(x)}\vert^2  $ stands for the variance. The pdf of a complex Gaussian distribution with mean $\mu$ and variance $\nu$ is represented by $\CN(x;\mu,\nu)$. The relation $f(x)=cg(x)$ for some positive constant $c$ is written as $f(x)\propto g(x)$.

\section{The New combined Message Passing Framework}
As a theoretical unified message passing framework, combined BP-MF algorithm~\cite{Riegler2013} classifies the factor nodes  $\mA$ on a factor graph into a BP subset $\mA_{\BP}$ and a MF subset $\mA_{\MF}$, fulfilling $\mA_{\BP}\cup \mA_{\MF}=\mA$ and $\mA_{\BP}\cap \mA_{\MF}=\emptyset$. Then the messages passed on the factor graph is updated by the following equations,
\begin{align}
m^{\text{BP}}_{f_a\to x_i}(x_i)&=\left\langle f_a(\x_a)\right\rangle_{\prod_{j\in N(a)\backslash i}n_{x_j\to f_a}(x_j)},\nonumber\\
&~~~~~~~~~~~~~~~~~~~~~~~~~~~~ a \in \mathcal{A}_{\mathrm{BP}}, i\in N(a)\label{eq:Rule1}\\
m^{\text{MF}}_{f_a\to x_i}(x_i)&=\exp\left\lbrace  \left\langle\log f_a(\x_a)\right\rangle_{\prod_{j\in N(a)\backslash i}n_{x_j\to f_a}(x_j)}\right\rbrace,\nonumber\\
&~~~~~~~~~~~~~~~~~~~~~~~~~~~~ a \in \mathcal{A}_{\mathrm{MF}},i\in N(a)\label{eq:Rule2}\\
n_{x_i\to f_a}(x_i)&\propto\prod_{b\in \mathcal{A}_{\text{BP}}(i)\backslash a}m^{\mathrm{BP}}_{f_b\to x_i}(x_i)
\prod_{c\in \mathcal{A}_{\mathrm{MF}}(i)}m^{\text{MF}}_{f_c\to x_i}(x_i),\nonumber\\
&~~~~~~~~~~~~~~~~~~~~~~~~~~~~ i \in \mathcal I, a\in N(i)\label{eq:Rule3}
\end{align}
where $N(a)$ and $N(i)$ denotes the subset of variable nodes neighboring the factor node $f_a$ and the subset of factor nodes neighboring the variable node $x_i$, respectively, $\mathcal{A}_{\text{BP}}(i)=\mathcal{A}_{\text{BP}}\cap N(i)$ and $\mathcal{A}_{\text{MF}}(i)=\mathcal{A}_{\text{MF}}\cap N(i)$.

\begin{figure}[!t]
\centering
\includegraphics[width=0.3\textwidth]{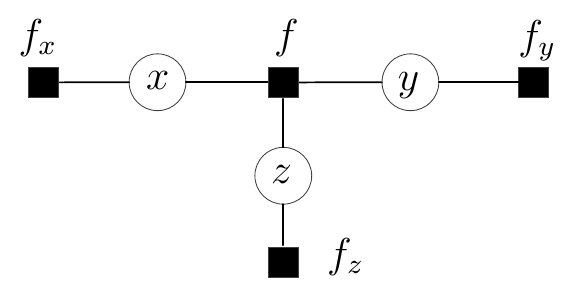}
\centering
\caption{A simple factor graph}
\label{fig:exam}
\end{figure}

The pure message passing algorithms, such as BP and MF, and the combined BP-MF only allow  one message update rule to handle a factor node. In this work, we heuristically propose a new combination, which allows one exploit both BP- and MF-like rule to deal with a single factor node, e.g. $f$ in Fig.~\ref{fig:exam}. To calculate the message from the factor node $f$ to variable node $y$, a BP-like rule is used at first,
\begin{equation}
\tilde{f}(x,y) = \left\langle f(x,y,z)\right\rangle_{n_{z \to f_\delta }(z)},
\end{equation}
then a MF-like rule is applied,
\begin{equation}
m_{f_\delta \to y}(y) = \exp\left\{\left\langle\log \tilde{f}(x,y)\right\rangle_{b(x)}\right\}.
\end{equation}
The new combination is well suitable for the case where the factor $f$ is a hard constraint, such as $f(x,y,z)=\delta(z-xy)$, and the message $n_{z \to f }(z)$ is of exponential form.
We will apply it to design a low complexity receiver for MIMO-OFDM systems in the following sections.

\section{System Model of  MIMO-OFDM Systems}
Consider the uplink of a multiuser MIMO-OFDM system which consists of a receiver equipped with $M$ antennas and $N$ users, each equipped with one antenna. To combat the inter-symbol interference, OFDM with $K$ subcarriers is adopted. The transmitted symbols by the $n$th user in frequency domain are denoted by $\bx_n=[x_n(1),...,x_n(K)]\tra$. Among the $K$ subcarriers, $K_p$ uniformly spaced subcarriers are selected for the users to transmit pilot signals, and the set of pilot-subcarriers of user $n$ is denoted by $\mP_n$. As in \cite{Wu2016}, we assumes that $\cap \mP_n=\emptyset$, and when a pilot-subcarrier is employed by a user, the remaining users do not transmit signals at the pilot-subcarrier. By the introduction of auxiliary variables $z_{mnk}=x_{nk}h_{mnk}$ and $\tau_{mk}=\sum_{n}z_{mnk}$, the received signal by the $m$th receive antenna at the $k$th subcarrier can be written as
\begin{eqnarray}
y_{mk}&=&\sum\nolimits_{n}{h_{mnk}x_{nk}}+\omega_{m}\nonumber\\
&=&\sum\nolimits_{n}{z_{mnk}}+\omega_{m}=\tau_{mk}+\omega_{m},\label{eq:recv_mod}
\end{eqnarray}
where $h_{mnk}$ stand for the frequency-domain channel weight between the $n$th transmit antenna and the $m$th receive antenna, and $\omega_{m}$ denotes the additive white Gaussian noise (AWGN) with zero mean and variance $\lambda^{-1}$. The transmitted and received symbol by the $n$th user and $m$th receiver in frequency domain are denoted by $x_{nk}$ and $y_{mk}$ respectively.

\begin{figure}[!t]
\centering
\includegraphics[width=0.48\textwidth]{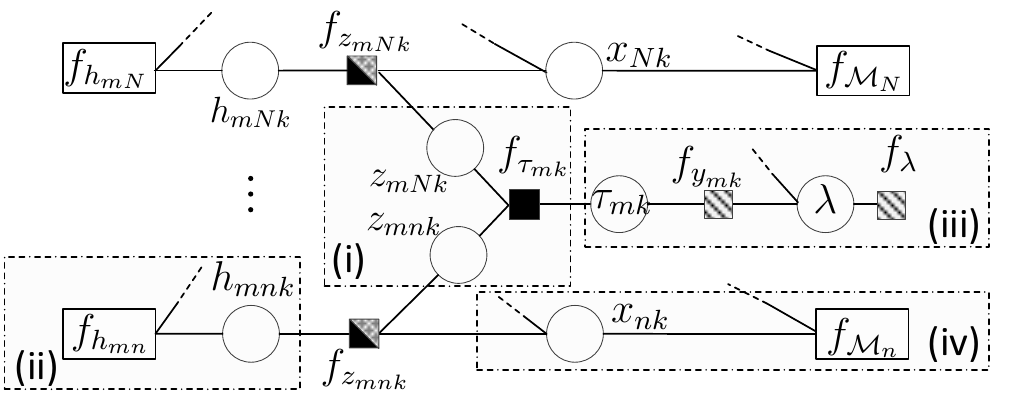}
\centering
\caption{A factor graph representation for the MIMO-OFDM system}
\label{fig:Model}
\end{figure}

From the receive model demonstrated in \eqref{eq:recv_mod}, the joint pdf of the collection of observed and unknown variables in the multi-signal model can be factorized as
\begin{eqnarray}
p(\y,\btau,\bz,\h,\x,\lambda)
=f_{\lambda}(\lambda)\prod_{m,n,k}f_{z_{mnk}}(z_{mnk},x_{nk},h_{mnk})\nonumber\\
\times \prod_{m,k} f_{\tau_{mk}}(\tau_{mk},\bz_{mk})f_{y_{mk}}(\lambda,\tau_{mk}) \prod_{m,n,k}f_{h_{mnk}} \prod_{n}f_{\mM_n},\label{eq:factorization}
\end{eqnarray}
where $f_{y_{mk}}(\lambda,\tau_{mk})\triangleq \CN(y_{mk};\tau_{mk},\lambda^{-1})$ denotes the observation node, factors $f_{\tau_{mk}}(\tau_{mk},\bz_{mk})=\delta(\tau_{mk}-\sum\nolimits_{n}{z_{mnk}})$ and $f_{z_{mnk}}(z_{mnk},x_{nk},h_{mnk})=\delta(z_{mnk}-x_{nk} h_{mnk})$ represents the constraint relationship of variables $z_{mnk},x_{nk},h_{mnk}$, and vector $\bz_{mk}$ denoted as $\bz_{mk}=[z_{m1k},...,z_{mNk}]\tra$. The modulation, coding, and interleaving constraints are denoted by $f_{\mM_n}$, and $f_{h_{mnk}}$ represents the priori of channel weight $h_{mnk}$. We assume that noise precision is unknown with the a priori $f_{\lambda}=1/\lambda$.
The above factorization lead to a factor graph representation shown in Fig.~\ref{fig:Model}.

\section{Receiver Design Using the New combined Message Passing Algorithm}
The difference between the proposed receiver and   that in \cite{Yuan2017} lies in how to calculate the messages related to the factor nodes $\{f_{z_{mnk}}\}$. \cite{Yuan2017} handles the factor nodes $\{f_{z_{mnk}}\}$ by using BP rule, and expectation propagation (EP) is also exploit to covert some messages to be Gaussian.
In this work,  we adopt the new combination method to deal with the factor nodes. Therefore,  the same messages in~\cite{Yuan2017} will not be listed, and we only detailed the computation of the messages $n_{x_{nk}\to f_{\mM_n}}(x_{nk})$ for soft demodulation, $n_{h_{mnk}\to f_{h_{mnk}}}(h_{mnk})$ for channel estimation and $m_{f_{z_{mnk}}\to z_{mnk}}(z_{mnk})$ for multi-signal interference cancellation.
In addition, we assume that the messages $m_{f_{h_{mn}}\to h_{mnk}}(h_{mnk})=\CN (h_{mnk};\vec{h}_{mnk},\vec{\nu}_{h_{mnk}})$, $m_{f_{\mM_n}\to x_{nk}}(x_{nk})=\sum\nolimits_{s}\gamma^s_{nk}\delta(x_{nk}-s)$ and $n_{ z{mnk}\to  f_{z_{mnk}}}(z_{mnk})=\CN(z_{mnk};{\cev{z}_{mnk},\cev{\nu}_{z_{mnk}}})$ are known, {and have listed in \cite{Yuan2017}}.
%

\subsection{The Message for Soft Demodulation}
At first, we apply a BP-like rule to the hard constraint factor node $f_{z_{mnk}}$, yielding
\begin{align}
&\tilde{f}_{z_{mnk}}(x_{nk},h_{mnk})\nonumber\\
&~~=\left\langle{f_{z_{mnk}}(z_{mnk},x_{nk},h_{mnk})}\right\rangle_{n_{z_{mnk}\to f_{z_{mnk}}}(z_{mnk})}\nonumber\\
&~~=\CN(x_{nk}h_{mnk};\cev{z}_{mnk},\cev{\nu}_{z_{mnk}}).
\end{align}
Then, the message $m_{f_{z_{mnk}}\to x_{nk}}(x_{nk})$ is updated by a MF-like rule,
\begin{align}
m_{f_{z_{mnk}}\to x_{nk}}(x_{nk})
&=\exp\left\{\left\langle \log \tilde{f}_{z_{mnk}}(x_{nk},h_{mnk})\right\rangle_{b(h_{mnk})}\right\}\nonumber\\
&\triangleq\CN\left(x_{nk};\vec{x}_{mnk},\vec{\nu}_{x_{mnk}}\right).\label{eq:z2x}
\end{align}
where $b(h_{mnk})=\CN(h_{mnk};\hat{h}_{mnk},\nu_{h_{mnk}})$ is presented later in \eqref{eq:blf_h}, and
\begin{eqnarray}
{\vec{x}_{mnk}}=\frac{\hat h^*_{mnk}\cev{z}_{mnk}}{|\hat h_{mnk}|^2+\nu_{h_{mnk}}},~~
{\vec{\nu}_{x_{mnk}}}=\frac{\cev{\nu}_{z_{mnk}}}{|\hat h_{mnk}|^2+\nu_{h_{mnk}}}.
\label{eq:msg_fz2x}
\end{eqnarray}

The message $n_{x_{nk}\to f_{\mM_n}}(x_{nk})$, passed to soft demodulation, is calculated by
\begin{eqnarray}
n_{x_{nk}\to f_{\mM_n}}(x_{nk})&=&\prod_{m,k}m_{f_{z_{mnk}}\to x_{nk}}(x_{nk})\nonumber\\
&\triangleq& {\CN(x_{nk};\hat\xi_{nk},\nu_{\xi_{nk}})}
\label{eq:msg_x2M}
\end{eqnarray}
where
\begin{eqnarray}
\nu_{\xi_{nk}}=\Big(\sum_{m}\frac{1}{\vec{\nu}_{x_{mnk}}}\Big)^{-1},~~~
\hat\xi_{nk}=\nu_{\xi_{nk}}\sum_{m}\frac{\vec{x}_{mnk}}{\vec{\nu}_{x_{mnk}}}.\nonumber
\end{eqnarray}

To update the message passed to channel estimation part, we have to  calculate the belief of $x_{nk}$, given as
\begin{eqnarray}
b(x_{nk})=\frac{ n_{x_{nk}\to f_{\mM_n}}(x_{nk})m_{f_{\mM_n}\to x_{nk}}(x_{nk})}{\int_{x_{nk}} n_{x_{nk}\to f_{\mM_n}}(x_{nk})m_{f_{\mM_n}\to x_{nk}}(x_{nk})}.\label{eq:blf_x}
\end{eqnarray}
Its mean and variance are
\begin{eqnarray}
\hat x_{nk}&=&\langle x_{nk}\rangle_{b(x_{nk})}\nonumber\\
\nu_{x_{nk}}&=&\langle |x_{nk}|^2 \rangle_{b(x_{nk})}-|\hat x_{nk}|^2.\nonumber
\end{eqnarray}

\subsection{The Message for Channel Estimation}
Similar to Eq. \eqref{eq:z2x}, the message $m_{f_{z_{mnk}\to h_{mnk}}}(h_{mnk})$ is also updated by a MF-like equation,
\begin{align}
&m_{f_{z_{mnk}\to h_{mnk}}}(h_{mnk})\nonumber\\
&~~~~~=\exp\left\{\left\langle\log \tilde{f}_{z_{mnk}}(x_{nk},h_{mnk})\right\rangle_{b(x_{nk})}\right\}\nonumber\\
&~~~~~\propto \CN\left(h_{mnk};\cev{h}_{mnk},\cev{\nu}_{h_{mnk}}\right)\label{eq:msg_fz2h}
\end{align}
where
\begin{align}
\cev{h}_{mnk}=\frac{\hat x_{nk}^*\cev{z}_{mnk}}{|\hat x_{nk}|^2+\nu_{x_{nk}}},~~~
\cev{\nu}_{h_{mnk}}=\frac{\cev{\nu}_{z_{mnk}}}{|\hat x_{nk}|^2+\nu_{x_{nk}}}.\nonumber
\end{align}
The message for channel estimation is $n_{h_{mnk}\to f_{h_{mk}}}(h_{mnk})=m_{f_{z_{mnk}\to h_{mnk}}}(h_{mnk}).$

The belief of $h_{mnk}$ is calculated as
\begin{eqnarray}
b(h_{mnk})&=&n_{h_{mnk}\to f_{h_{mnk}}}(h_{mnk})m_{f_{h_{mnk}}\to h_{mnk}}(h_{mnk})\nonumber\\
&\triangleq&\CN\left(h_{mnk};\hat h_{mnk},\nu_{h_{mnk}}\right)\label{eq:blf_h}
\end{eqnarray}
where
\begin{align}
&\nu_{h_{mnk}}=\left(\cev{\nu}_{h_{mnk}}^{-1}+\vec{\nu}_{h_{mnk}}^{-1}\right)^{-1}\label{eq:blf_h_v}\\
&\hat h_{mnk}=\nu_{h_{mnk}}(\vec{h}_{mnk}/\vec{\nu}_{h_{mnk}}+\cev{h}_{mnk}/\cev{\nu}_{h_{mnk}}).
\label{eq:blf_h_m}
\end{align}

\subsection{The Message for Multi-signal Interference Elimination}
Since the factor node $f_{z_{mnk}}$ stands for the hard constraint  $z_{mnk}=x_{nk}h_{mnk}$,  the belief of $b(z_{mnk})$ can be equivalently computed as
\begin{align}
b(z_{mnk})=\left\langle f_{z_{mnk}}(z_{mnk},x_{nk},h_{mnk})\right\rangle_{b(x_{nk})b(h_{mnk})}\nonumber
\end{align}
and its mean and variance are easily obtained
\begin{align}
&\hat z_{mnk}=\hat x_{nk} \hat h_{mnk} \label{eq:blf_z_m}\\
&\nu_{z_{mnk}}=|\hat x_{nk}|^2\nu_{h_{mnk}}+|\hat h_{mnk}|^2\nu_{x_{nk}}+\nu_{h_{mnk}}\nu_{x_{nk}}.\label{eq:blf_z_v_Mix}
\end{align}

Then, the message $m_{f_{z_{mnk}}\to z_{mnk}}(z_{mnk})$ is calculated as~\footnote{For a pdf $b(x)$, $\Proj\{b(x)\}=\CN(x;m,\nu)$, where $m=\langle x\rangle_{b(x)}$ and $\nu =\langle |x|^2\rangle_{b(x)}-|m|^2$, stands for projecting a function $b(x)$ to Gaussian family.}
\begin{eqnarray}
m_{f_{z_{mnk}}\to z_{mnk}}(z_{mnk})&=&\frac{\Proj\left\{ b(z_{mnk})\right\}}{n_{z_{mnk}\to f_{z_{mnk}}}(z_{mnk})}\nonumber\\
&\triangleq& \CN\left(z_{mnk};\vec{z}_{mnk},\vec{\nu}_{z_{mnk}}\right),\nonumber
\end{eqnarray}
where
\begin{align}
&\vec{\nu}_{{z}_{mnk}}=\left({1}/{\nu_{z_{mnk}}}-{1}/{\cev{\nu}_{{z}_{mnk}}}\right)^{-1}
\label{eq:msg_z2ftau_m}\\
&\vec{z}_{mnk}=\vec{\nu}_{{z}_{mnk}}\left({\hat z_{mnk}}/{\nu_{z_{mnk}}}
-{\cev{z}_{mnk}}/{\cev{\nu}_{{z}_{mnk}}}\right).\label{eq:msg_z2ftau_v}
\end{align}

\section{Simulation Results and Complexity Comparison}\label{Sec:sim}
Consider a MIMO-OFDM system with the same parameters as in \cite{Yuan2017}. We compare the proposed receiver with four state-of-the-art receivers in the literature\footnote{For a fair comparison and explicitly demonstrating the virtues of the proposed receiver, all considered receivers update message $m_{f_{\bh_{mn}}\to h_{mnk}}(h_{mnk})$ using the method in \cite{Yuan2017}.}: (1)Aux: the auxiliary variable aided method proposed in \cite{Yuan2017}. (2)BP-MF: disjoint version of the receivers in \cite{Navarro2011}; (3) {BP-MF-EPv}: the receiver proposed in \cite{Daniel2016}; (4) {BP-EP-GA}: the BP-EP-based receiver in~\cite{Wu2016} with perfect noise precision. As a reference, the performance of the receiver with perfect channel weight $\bh$, noise precision $\lambda$ and multiuser interference cancellation is also included, denoted by matched filter bound (MFB).

\begin{figure}[!t]
\centering
\includegraphics[width=0.45\textwidth]{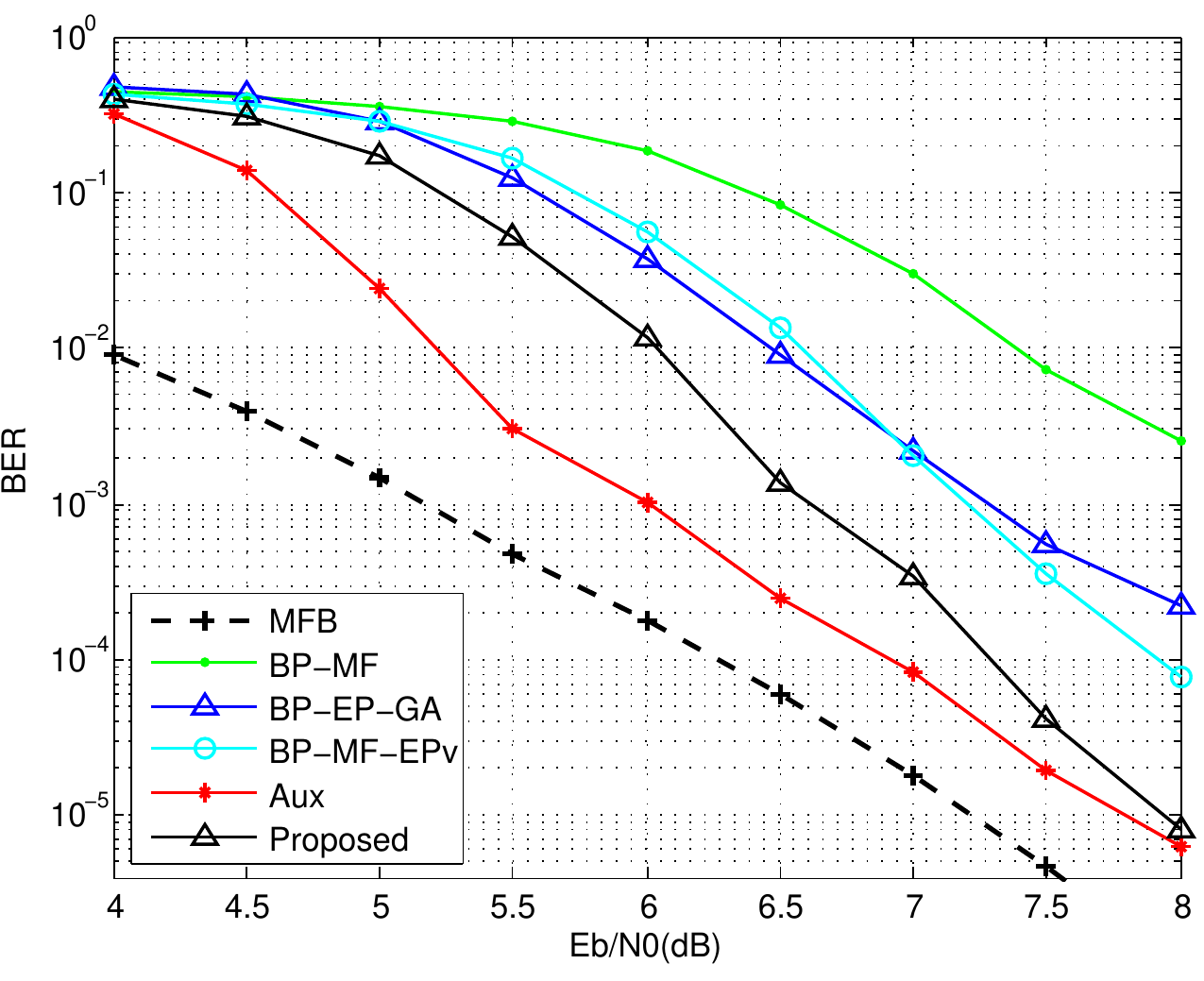}
\caption{BER performance of different algorithms.}\label{fig:BERvsSNR}
\end{figure}

Fig.~\ref{fig:BERvsSNR} shows the BER performance of the different receivers with running 15 iterations. The proposed low complexity receiver has a performance loss of 0.5dB compared to the high complexity version, ``Aux", in the moderate Eb/N0.  Meanwhile, it achieves a performance gain of more than 1dB compared to ``BP-MF", and outperforms ``BP-EP-GA" and ``BP-MF-EPv" by about 0.5dB. Especially, the perform of  proposed receiver can approach that of ``Aux" at higher SNRs.

Since all the receivers employed the same method in the updating of messages $\{m_{f_{\mM_n}\to x_{nk}}(x_{nk})\}$ and $\{n_{h_{mnk}\to {f_{z_{mnk}}}}(h_{mnk})\}$, here we only compare the complexity in handling the multiplication and multi-signal summation problem. The complexity of the proposed receiver, ``BP-MF" and ``BP-EP-GA" is in the order $\mO(MNK+NKQ)$, since $MNK$ Gaussian message and $NK$ discrete beliefs should be calculated, while the ``BP-MF-EPv" has complexity of $\mO(MN^2K+KN^3+NKQ)$, where $Q$ is the modulation order. Since $MNK$ Gaussian mixture belief should be computed, the complexity of ``Aux" is  $\mO(MNKQ)$.

\section{Conclusion}
In this paper, we propose a new combined message passing framework which will lead more flexible combination of BP and MF on factor graphs. It is applied to design a low complexity receiver for MIMO-OFDM systems.
The receiver using the proposed message passing algorithm can obtain better trade-off between performance and complexity that the state-of-the-art receivers.
\bibliographystyle{IEEEtran}
\bibliography{IEEEabrv,bibliography}

\end{document}

%% file: mydefinitions.tex
\newcommand{\y}{\boldsymbol{y}}
\newcommand{\x}{\boldsymbol{x}}
\newcommand{\h}{\boldsymbol{h}}

\newcommand{\bx}{\boldsymbol{x}}

\newcommand{\bh}{\boldsymbol{h}}

\newcommand{\bz}{\boldsymbol{z}}

\newcommand{\btau}{\boldsymbol{\tau}}

\newcommand{\mP}{\mathcal{P}}
\newcommand{\mG}{\mathcal{G}}
\newcommand{\mO}{\mathcal{O}}
\newcommand{\mA}{\mathcal{A}}

\newcommand{\mM}{\mathcal{M}}

\newcommand{\cev}[1]{\reflectbox{\ensuremath{\vec{\reflectbox{\ensuremath{#1}}}}}}

\newcommand{\MF}{\mathrm{MF}}
\newcommand{\BP}{\mathrm{BP}}

\newcommand{\Proj}{\textrm{Proj}_{\mG}}

\newcommand{\tra}{^{\textrm{T}}}

\newcommand{\CN}{\mathcal{CN}}